\documentclass[superscriptaddress,aps,prl,showkeys,showpacs,twocolumn,10pt]{revtex4}
\usepackage[OT1,T1]{fontenc}
\usepackage{graphicx}
\usepackage{rotating}
\begin{document}

\title{ Examination of the Nature of the ABC Effect}
\date{\today}

\newcommand*{\PITue}{Physikalisches Institut, Eberhard--Karls--Universit\"at 
  T\"ubingen,  Auf der Morgenstelle~14, D-72076 T\"ubingen, Germany}
\newcommand*{\Kepler}{Kepler Center for Astro and Particle Physics, University
  of T\"ubingen,  Auf der Morgenstelle~14, D-72076 T\"ubingen, Germany}
\newcommand*{\Edin}{School of Physics and Astronomy, University of Edinburgh,
James Clerk Maxwell Building, Peter Guthrie Tait Road, Edinburgh, EH9 3FD, UK}
\newcommand*{\Tomsk}{Department of Physics, Tomsk State University, 36 Lenina
  Avenue, Tomsk 634050, Russia}

\author{M.~Bashkanov\footnote{corresponding author\\ email address: mikhail.bashkanov@ed.ac.uk}}   \affiliation{\Edin} 
\author{H.~Clement}     \affiliation{\PITue},\affiliation{\Kepler}
\author{T.~Skorodko}     \affiliation{\PITue},\affiliation{\Kepler},\affiliation{\Tomsk}

\begin{abstract}
Recently it has been shown by exclusive and kinematically complete experiments
that
the appearance of a narrow resonance structure in double-pionic fusion
reactions is strictly correlated with the appearance of the so-called ABC
effect, which denotes a pronounced low-mass enhancement in the
$\pi\pi$-invariant mass spectrum. Whereas the resonance structure got its
explanation by the $d^*(2380)$ dibaryonic resonance, a satisfactory
explanation for the ABC effect is still pending. In this paper we discuss
possible explanations of the ABC effect and their consequences for the 
internal structure of the $d^*$ dibaryon. To this end we examine and review a
variety of 
proposed explanations for the ABC effect, add a new hypothesis and
confront all of them with the experimental results for the $np \to d\pi^0\pi^0$
and $np \to np\pi^0\pi^0$ reactions, which are the most challenging ones for
this topic. 
\end{abstract}

\maketitle

\section{Introduction}

Recently it was 
shown 
that there is a resonance pole at $(2380\pm10) - i(40\pm5)$ MeV in the $^3D_3
-^3G_3$ coupled partial waves
based on a SAID partial-wave analysis, which included new data from COSY on the
analyzing power of $np$ scattering 
\cite{prl2014,npfull,RW,RWnew}. 
This finding matches perfectly with the $I(J^P) =
0(3^+)$ resonance structure observed at $\sqrt s \approx$ 2.37 GeV with a
width of 70 MeV in the total cross section of the double-pionic fusion reactions
$pn \to d\pi^0\pi^0$ and $pn \to d\pi^+\pi^-$
\cite{prl2009,prl2011,isofus}. 
The results from the WASA-at-COSY for the $pn \to d\pi^+\pi^-$ reaction have
meanwhile found support by preliminary results from HADES \cite{Kuc}. 

Having revealed the pole in the $np$ scattering
amplitudes 
means that this resonance structure constitutes an 
$s$-channel resonance in the system of two baryons. It has been denoted since
then by $d^*(2380)$ following the nomenclature used for nucleon excitations. 

Follow-up measurements of the non-fusion reactions $pn \to
pp\pi^0\pi^-$ \cite{pp0-} and $pn \to pn \pi^0\pi^0$ \cite{np00} with the WASA
detector at COSY and $np \to np\pi^+\pi^-$ \cite{hades} with the HADES
  detector at GSI showed that also these reactions, which are partially of 
isoscalar character, show the $d^*(2380)$ resonance in agreement with
expectations based on isospin coupling. 

In addition, WASA measurements revealed that $d^*(2380)$ is also present in
the double-pionic fusion reactions to the helium isotopes $pd \to
^3$He$\pi^0\pi^0$, $pd \to ^3$He$\pi^+\pi^-$, $dd \to ^4$He$\pi^0\pi^0$ and
$dd \to ^4$He$\pi^+\pi^-$ \cite{3he,MB,4he,SK}. This means that obviously
$d^*(2380)$ is stable enough to survive also in a nuclear surrounding. This
conclusion is 
     in agreement with 
the appearance of a dilepton enhancement (DLS
puzzle) in heavy-ion collisions \cite{DLS}.   

\section{ABC effect}

In 1960 Abashian, Booth and Crowe \cite{abc} found out that the
$\pi\pi$-invariant mass spectrum in the double-pionic fusion reaction 
$pd \to ^3$He$\pi\pi$ exhibits a pronounced low-mass enhancement. Subsequent
measurements showed that this enhancement also persists in the fusion reactions
to $d$ and $^4$He \cite{bar,abd,hom,hal,ba,ban,ban1,plo,col,wur,cod}, if
the produced pion pair is of isoscalar nature. Since 
there was no plausible explanation for this effect, it got named after the
initials of the authors of the first publication on that issue.

The recent exclusive and kinematically complete WASA measurements, which were
carried out at CELSIUS and later-on at COSY, confirmed the previous 
findings, which had been obtained by merely inclusive single-arm
magnetic spectrometer and low-statistics bubble-chamber measurements,
respectively. Fig.~1 shows as an example the $\pi^0\pi^0$-invariant mass
spectrum obtained in the reaction $pn \to d\pi^0\pi^0$ at $\sqrt s$ = 2.38
GeV \cite{prl2011}. Relative to a pure phase-space distribution, the data
exhibit a very pronounced low-mass enhancement -- the ABC effect.

\begin{figure}
\centering
\includegraphics[width=0.99\columnwidth]{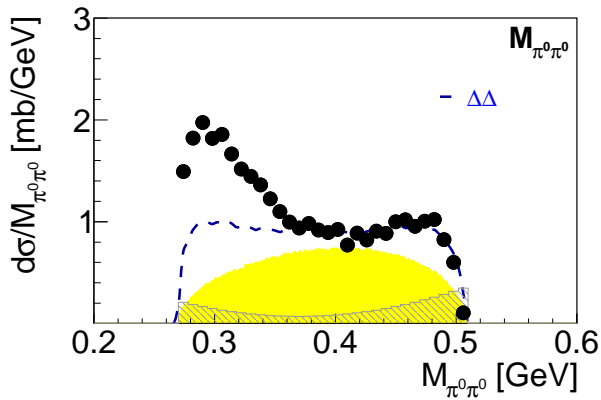}
\caption{\small (color online). $\pi^0\pi^0$ invariant mass distribution in
  the $pn \to d\pi^0\pi^0$ reaction at $\sqrt s$ = 2.38 GeV, the $d^*(2380)$
  resonance peak energy region. Black solid circles represent the data and the
  hatched area an estimate of systematic uncertainties (from
  Ref. \cite{prl2011}). The yellow shaded area denotes pure phase space and
  the dashed line gives a calculation of the conventional $t$-channel
  $\Delta\Delta$ excitation process $pn \to \Delta\Delta \to d\pi^0\pi^0$,
  both normalized arbitrarily in height.  
}
\label{fig1}
\end{figure}

Moreover, the new measurements discovered a strict correlation
between the 
appearance of the ABC effect and the appearance of the $d^*(2380)$ resonance
structure in the total cross sections of the fusion reactions to d,
$^3$He and $^4$He, if these reactions were associated with the production of
an isoscalar pion pair \cite{prl2009,prl2011,isofus,3he,4he}.
 
In contrast to these findings no ABC effect was observed in the non-fusion
reactions $pn \to pp\pi^0\pi^-$ and $pn \to pn\pi^0\pi^0$ despite of the
appearance of $d^*(2380)$ in these reactions. For the first reaction the
non-appearance of the ABC effect is easily understood, since in this case the
pions of the isovector pion pair must be in relative $p$-wave suppressing thus
any low-mass enhancement. For the second reaction there is no such obvious
reason.  

In the following we examine possible scenarios as suitable explanations of the
ABC effect. To this end we confront several phenomenological ansatzes with the
observations made -- with special emphasis on the $np
\to d\pi^0\pi^0$ and $np \to np\pi^0\pi^0$  reactions, where the largest
tensions for a common explanation exist. Hence the work presented in the
following is intended to be primarily a phenomenological search for a
reasonable solution rather than presenting here already a 
fundamental theoretical treatment of this issue, which would need to be a next
step as soon as a reasonable phenomenological solution has been found.
  
The aim of this paper is also to document calculations of the $pn \to d^* \to
\Delta\Delta$ route, the results of which were shown in previous publications
\cite{prl2011,isofus,pp0-,np00}.

\section{Hypotheses for its Explanation}

From the Dalitz plots of the double-pionic fusion reactions
\cite{prl2009,prl2011,isofus,3he,4he} we know that $d^*(2380)$ decays 
predominantly via the $\Delta\Delta$ system in the intermediate state -- with
the two $\Delta$s being in relative $s$-wave. This is in accordance with
meanwhile numerous theoretical work about this resonance
\cite{Dyson,kamae,ping,ping1,GG1,GG2,shen,zhang1,zhang2,chen,dong}.  It also
is in agreement with the measured branchings of the $d^*$ decay into the
diverse $NN\pi\pi$ channels \cite{BCS}.

Therefore it seems likely that the 
appearance of the ABC effect is correlated with the appearance of the
$\Delta\Delta$ system in the intermediate state and that way also 
with the internal structure of $d^*(2380)$.

\begin{figure}
\centering
\includegraphics[width=0.7\columnwidth]{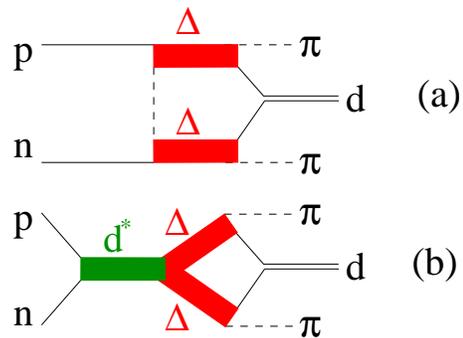}
\caption{\small (color online). Schematic representation of the processes $pn \to d\pi\pi$ via
  $t$-channel meson exchange (a) and $pn \to d^*(2380) \to
  \Delta\Delta \to d\pi\pi$ (b).
}
\label{fig2}
\end{figure}

One of the first attempts to explain the ABC effect was its connection
with the conventional $t$-channel $\Delta\Delta$ excitation, a schematic
diagram of which is depicted in Fig.~\ref{fig2}(a). As pointed out already by Risser
and Shuster \cite{ris}, such an excitation causes a double-hump 
structure of the $\pi\pi$-invariant mass spectrum with enhancements both at
low and high masses relative to phase space. If the center-of-mass energy
$\sqrt s$ is below the mass 
$2m_\Delta$, then the low-mass enhancement tends to be more prominent -- see
Fig.~3. If it is above 
$2m_\Delta$, then the high-mass enhancement dominates. 

Fig.~\ref{fig1} exhibits the $\pi^0\pi^0$-invariant mass ($M_{\pi^0\pi^0}$)
spectrum  at
resonance in the $pn \to d\pi^0\pi^0$ reaction, where the background from
conventional processes is small compared to the resonance signal
and can be neglected in the following. 
The dashed curve shows the result, if we calculate the $t$-channel
$\Delta\Delta$ process (Fig.~2(a)) as given by Risser and Shuster \cite{ris}.

\begin{figure}
\centering
\includegraphics[width=0.99\columnwidth]{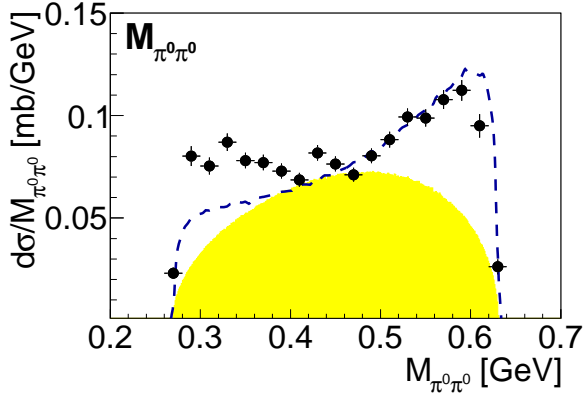}
\caption{\small (color online). The same as Fig.~1, but at $\sqrt s$ = 2.5
  GeV, {\it i.e.}, above the $d^*(2380)$ region in the regime of the
  $t$-channel $\Delta\Delta$ process. The data are from Ref. \cite{prl2011}.
} 
\label{fig1a}
\end{figure}

For comparison we show in Fig.~\ref{fig1a} data \cite{prl2011} for the
$M_{\pi^0\pi^0}$ spectrum at $\sqrt s$ = 2.5 GeV, {\it i.e.}, at an energy
above the $d^*(2380)$ region in the regime of the conventional $t$-channel
$\Delta\Delta$ process. The data are accounted reasonably well by the
calculations in the framework of Risser and Shuster. Since
$\sqrt s > 2m_\Delta$ here, the high-mass enhancement is now
dominating and correctly reproduced by these calculations. To the observed
low-mass enhancement also the high-energy tail of $d^*(2380)$ may contribute
-- as well as possibly a $\Delta\Delta$ final-state interaction as discussed
below.

\subsection{Description of the $d^*(2380)$ Resonance Process}

In the following we give the formalism for calculating the 
 $s$-channel resonance process 

\begin{equation}
  pn \to d^*(2380) \to \Delta\Delta \to d\pi^0\pi^0
\end{equation}

depicted schematically in
Fig.~\ref{fig2}(b) and as used for the calculations presented in
Ref.~\cite{prl2011,isofus,pp0-,np00}.

The differential cross section is then given by 
\begin{equation}
d\sigma = \frac{(2\pi)^4}{4~p_i~\sqrt{s}}~\sum |{\it A(m_p,m_n,m_d,{\hat
    k_1},{\hat k_2}) 
}|^2 ~d\Phi_3,
\end{equation}
where $d\Phi_3$ denotes the 3-body phase-space element and the sum runs over
the observable spin substates $m_p,m_n,m_d$ of proton, neutron and deuteron, as
well as over all kinematical variables not considered explicitly in the
dependence of the differential cross section $d\sigma$. $\vec k_1$ and
$\vec k_2$ denote the momentum vectors of the emitted pions and $p_i$ is the
momentum in the initial $pn$ channel.

The transition amplitude ${\it A(m_p,m_n,m_d,{\hat k_1},{\hat k_2})}$ is
given by  

\begin{equation}
{\it A(m_p,m_n,m_d,{\hat k_1},{\hat k_2})} \sim {\int M_R * {\hat \varphi_d}~d{\vec q}},
\end{equation}

where $q$ denotes the relative momentum between the two nucleons merging
finally into the deuteron and
${\hat \varphi_d}$ denotes the deuteron wave function in momentum space. Its
square provides the probability distribution for finding the two involved
nucleons 
with momenta $p_{N_1}$ and $p_{N_2}$. 
    In our numerical calculations we used $q_{max}$ = 800 MeV/c as an upper
    limit for the relative momentum $q$. 
    Extending the upper limit to 1.5 GeV/c yields no significant changes in the
    calculated observables, which would be of relevance for the considerations
    in the following.
Since we are interested here only in the
shape of the $\pi\pi$ invariant mass distribution, we omit constants and use
the proportionality sign in eq. (3).

The matrix element $M_R$ is given by
\begin{equation}
M_R(m_p,m_n,m_d,{\hat k_1},{\hat k_2}) = M_R^0 ~~\Theta_R(m_p,m_n,m_d,{\hat
    k_1},{\hat k_2}),
\end{equation}
where the function $\Theta$ contains the substate and angular dependent part,
and 
\begin{equation}
M_R^0 =  D_R * D_{\Delta_1} *
D_{\Delta_2}.
\end{equation}
Here $D_{\Delta}$ denotes the $\Delta$ propagator 
\begin{equation}
D_{\Delta} =  \frac{\sqrt{m_{\Delta}
      ~\Gamma_{\Delta}~/~q_{\pi}^{N\pi}}}{M_{N\pi}^2 - m_{\Delta}^2 + i
    m_{\Delta} \Gamma_{\Delta}}
\end{equation}
and 
\begin{equation}
\Gamma_{\Delta}(q_{\pi}^{N\pi}) = \gamma (q_{\pi}^{N\pi})^3 \frac{R^2}{1 +
  R^2 (q_{\pi}^{N\pi})^2}
\end{equation}
with $\gamma = 0.74$, $R = 6.3 (GeV/c)^{-1}$ and $q_{\pi}^{N\pi}$ denoting the
decay momentum of the pion in the $\Delta$ system \cite{ris}. $M_{N\pi}$
refers to the invariant mass of the $N\pi$ system.

If we write $D_R$ in form of a Breit-Wigner amplitude, then we have 
\begin{equation}
M_R^0 =             
\frac{m_R\sqrt{f~\Gamma_i ~\Gamma_{\Delta\Delta}}}{s -
    m_R^2 + i m_R \Gamma_R} ~ D_{\Delta_1} ~D_{\Delta_2}
\end{equation}
with mass $m_R \approx$ 2.38 GeV and width $\Gamma_R(s = m_R^2) \approx$ 70~MeV. \\

The factor 
\begin{equation}
f = \frac{s}{p_i~p_f} \frac{2J + 1}{(2s_p + 1) (2s_n + 1)}
\end{equation}
is the flux factor of the Breit-Wigner resonance. Here 
$p_f = p_{\Delta\Delta}$, given in eq.(11), is the 
decay momentum of each $\Delta$ in the $d^*$ system
 and $s_p$ and $s_n$ denote the nucleon spins
in the initial channel.
 
$\Gamma_i$ and $\Gamma_{\Delta\Delta}$ denote the partial widths
for formation and decay of the {\it s}-channel resonance. They are momentum
dependent quantities and proportional
to the square of the respective (effective) coupling constants $g_{pn}$ and
$g_{\Delta\Delta}$.

In case of $d^*$ with $I(J^P) = 0(3^+)$ the resonance is created by the 
$^3D_3 - ^3G_3$ coupled partial waves between $p$ and $n$ in the initial
system \cite{prl2014,npfull,RW}. Hence we have $\Gamma_i = g_{pn}^2
p_i^{2L+1}$, where $L$ denotes the orbital angular momentum in the initial
$pn$ channel.  
Since the resonance is
far beyond the $pn$ threshold, the momentum dependence over the comparatively
narrow region of the resonance is small and can actually be neglected here. 

In the exit channel the $d^*$ resonance decays into the $\Delta\Delta$ system
with a relative $S$-wave between the two $\Delta$s --- as observed in the $\Delta$
angular distribution (Fig. 5 in Ref. \cite{prl2011}). Therefore we have 
\begin{equation}
\Gamma_{\Delta\Delta} = g_{\Delta\Delta}^2 p_{\Delta\Delta},
\end{equation}
with
\begin{eqnarray}
2\vec p_{\Delta\Delta} &&= \vec p_{\Delta_1} - \vec p_{\Delta_2} \\\nonumber
&&= (\vec p_{N_1} + \vec k_1) -  (\vec p_{N_2} +
\vec k_2).\nonumber
\end{eqnarray}

   Note that the $p_{\Delta\Delta}$ dependences in eqs. (9) and (10) cancel
   each other, when inserted into eq. (8).

The momentum dependent total width of the resonance is then given by 
\begin{eqnarray}
&&\Gamma_R(s) = \Gamma_i + \sum \Gamma_f = \Gamma_i +  \gamma_R *\\ 
&&\int dm_1^2 dm_2^2 p_{\Delta\Delta}
| D_{\Delta_1}(m_1^2) D_{\Delta_2}(m_2^2)|^2 \theta(s-(m_1+m_2)^2)\nonumber,
\end{eqnarray}
where 
$m_1^2$  and $m_2^2$ denote the invariant
mass-squared of the $N\pi$ pairs forming the systems $\Delta_1$ and
$\Delta_2$, respectively \cite{CH}. 
The integral runs then over squared invariant masses from the pion emission
threshold, given by $m_i^2 = (m_N + m_\pi)^2$ in the $NN\pi\pi$ channels,
to upper bounds defined by the $\theta$ function, which
ensures energy conservation. 


The second term in eq. (12) denotes the decays of the resonance via the
intermediate $\Delta\Delta$ system. The quantity $\gamma_R$  contains the
coupling constant $g_{\Delta\Delta}$ and other constants and is fitted to
yield a total width of $\Gamma_R(s = m_R^2)$ = 70 MeV. 
In the following sections
we discuss also proposed scenarios, where the intermediate state is not only
$\Delta\Delta$. In such cases eq. (12) has to be complemented accordingly.

The decay of
$d^*(2380)$ into its various $NN\pi\pi$ channels and into the $np$ channel
have been 
discussed in detail in Ref. \cite{BCS}. There also the experimentally
extracted partial widths $\Gamma_i(s = m_R^2)$ and $\Gamma_f(s = m_R^2)$ are
given and can be deduced, respectively, from the quoted partial cross sections
at resonance.

The angular dependence is obtained from the angular momentum coupling in
entrance and exit channels: 
\begin{equation}
\vec s_p + \vec s_n + \vec {L} = \vec {J} = \vec s_{\Delta_1} + \vec
s_{\Delta_2}
\end{equation}
with 
\begin{eqnarray}
\vec {s}_p + \vec {s}_n =&& \vec {s}, ~~~\\\nonumber
\vec s_{\Delta_1}=&& \vec l_1 + \vec s_{N_1}, ~~~\\\nonumber
\vec s_{\Delta_2} =&& \vec l_2 + \vec s_{N_2}, ~~~\\\nonumber
\vec s_{N_1} + \vec s_{N_2} =&& \vec s_{d}. \nonumber
\end{eqnarray}
Here $\vec s_p$, $\vec s_n$ and $\vec s_d$ with $s_p = s_n = \frac {1}
{2}$ and $s_d = 1$ denote the spins of $p, n$
and $d$. The orbital angular momenta $\vec l_1$ and $\vec l_2$ with $l_1 =
l_2 = 1$ stand for the pion $p$-waves originating from $\Delta$ decay and $\vec
{L}$ is the initial orbital angular momentum between the incident proton and
neutron. 
Since we deal here with a resonance in the $^3D_3$-$^3G_3$ coupled
partial waves the spin of the nucleon pair in the initial channel is $s$ = 1.

Eq. (13) assumes that we have $s$-wave between the two $\Delta$s in the
intermediate state -- in agreement with observation, as mentioned already above. 

If the coordinate system is chosen to be the standard one with the z-axis
pointing in beam direction (implying $m_L$ = 0 and $(\Theta_i, \Phi_i) =
(0,0)$),  
then the function $\Theta_R(m_p,m_n,m_d,{\hat k_1},{\hat k_2})$ defined in
eq. (3) is built up by the corresponding vector coupling coefficients
  and spherical harmonics representing the angular dependence due to the orbital
  angular momenta involved in the reaction: 
\begin{eqnarray}
\Theta_R(m_p,m_n,m_d,&&{\hat k_1},{\hat k_2}) =\\ \nonumber
&& \sum (\frac {1}{2} \frac {1}{2}
m_p m_n | 1 m_s) ~~ (1 L m_s0 | J M) ~~\\ \nonumber
&&(J M | \frac {3}{2} \frac {3}{2} m_1^{\Delta} m_2^{\Delta}) ~~
(\frac {3}{2} m_1^{\Delta} | \frac {1}{2} 1 m_1^N m_1) \\\nonumber
&&(\frac {3}{2} m_2^{\Delta} | \frac {1}{2} 1 m_2^N m_2) ~~
(1 m_d | \frac {1}{2} \frac {1}{2} m_1^N m_2^N)\\
\nonumber   &&Y_{L0}(0,0)
~~~Y_{1 m_1}(\hat {k_1}) ~~Y_{1 m_2}(\hat {k_2}).
\end{eqnarray}

The angular distributions for deuteron and pions resulting from  
eq. (15) are displayed in Fig. 5 of Ref.\cite{prl2011}. 

The $p$-wave pions emerging from the intermediate $\Delta\Delta$ system can
couple to relative $s$- and $d$-waves. In the first instance, $d$ and $\pi\pi$
systems must then be in relative $d$-wave to match the requirement for the
resonance spin, whereas in the second case $d$ and $\pi\pi$ systems have to be
in relative $s$-wave.

The calculation of the observables resulting from the process in
Fig.~\ref{fig2}(b) has 
been realized by use of the Monte-Carlo technique. There, 
after generation of the deuteron momentum within the reaction phase space the
momenta of the nucleons inside the deuteron are diced to be distributed
according to the deuteron wave function. By subsequent use of the formulae
given in eqs. (2 - 15) the calculation then conforms to the process
represented by the diagram in Fig.~\ref{fig2}(b). 
    The boson symmetrization has been realized by proper reshuffling of the
    Monte-Carlo events.  
\footnote{We note that this
  technique allows to be applied straightforwardly also for the $^3$He and
  $^4$He cases by inclusion of the nucleons' motion according to the
  respective wavefunctions -- see Refs. \cite{3he,4he}.}. The correctness of
the Monte-Carlo results has been cross-checked by respective analytic
calculations of the reaction amplitude corresponding to the diagram in
Fig.~\ref{fig2}(b) 
\cite{CH}.

\begin{figure}
\centering
\includegraphics[width=0.99\columnwidth]{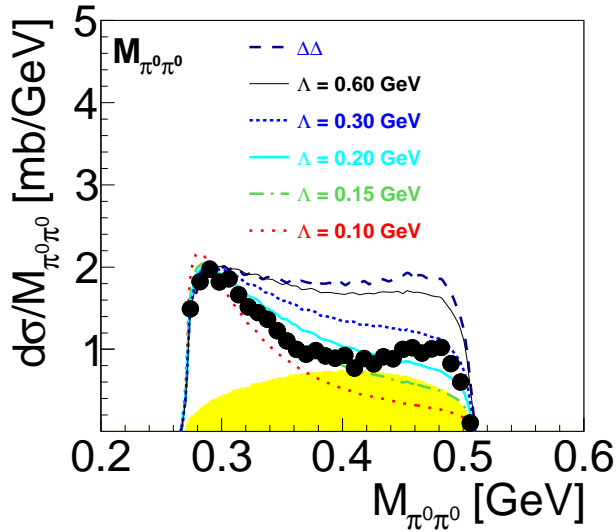}
\caption{\small (color online). Illustration of the vertex
  function effect on the $\pi^0\pi^0$ invariant mass distribution of the $pn
  \to d\pi^0\pi^0$ reaction at $\sqrt s$ = 2.38 GeV. Black solid
  circles give the data from Ref. \cite{prl2011}, the yellow shaded area
  denotes pure phase space (arbitrarily normalized). The dashed line
  represents the calculation of the process $pn \to d^*(2380) \to \Delta\Delta
  \to d\pi^0\pi^0$ without the vertex function from eq. (16). The other
  lines give the calculations with vertex function for cutoff parameters
  $\Lambda$ = 0.1, 0.15, 0.2, 0.3 and 0.6 GeV/c. The curves have been
  normalized to the data point at $M_{\pi^0\pi^0}$ = 0.29 GeV.
}
\label{fig3}
\end{figure}

For the $\pi^0\pi^0$-invariant mass spectrum the calculation of this
resonance process is shown by the dashed line in Fig.~\ref{fig3}. It gives the
proper 
tendency. However, the produced low-mass enhancement is much too small in
comparison to that observed in the data.

In fact, the calculation for the $M_{\pi^0\pi^0}$ distribution is practically
identical to that of the $t$-channel process  depicted in Fig.~\ref{fig2}(a)
and shown 
by the dashed curve in Fig.~\ref{fig1}. This is not surprising,
since at the energy of interest we are much below the nominal $\Delta\Delta$
threshold of $2m_\Delta$, so that also in this case the intermediate
$\Delta\Delta$ system is strongly favored to be in relative $s$-wave. Hence
the dynamics of the $\pi^0\pi^0$ system observed in the  $M_{\pi^0\pi^0}$
spectrum appears to be similar.
And since the $t$-channel $\Delta\Delta$ process is the dominant background for
the $d^*$ formation process, any interference between both processes will not
affect the shape of the $M_{\pi^0\pi^0}$ spectrum.

\subsection{$d^*\Delta\Delta$ Vertex  Function}

In order to heal the deficiency in the calculated low-mass enhancement, a
vertex function at the $d^*\Delta\Delta$ vertex  was introduced in
Ref.~\cite{prl2011} to account for finite size effects. It multiplies the
expression in eq.~(10) and was chosen to be of
monopole type 

\begin{equation}
FF \sim \frac{\Lambda^2}{\Lambda^2+p_{\Delta\Delta}^2}.
\end{equation}


If we neglect the Fermi motion of the nucleons within the deuteron, then the
nucleon momenta cancel in eq. (11) and we have $p_{\Delta\Delta} =
p_{\pi^0\pi^0}$, where $p_{\pi^0\pi^0}$ is the 
relative pion momentum. 
This means that the $p_{\Delta\Delta}$ dependence of the $d^*$ resonance term
in eq. (8) is directly reflected in the  $M_{\pi^0\pi^0}$ spectrum.  And since
the $p_{\Delta\Delta}$ dependence 
in eq. (8) solely rests in
the vertex function, we have here the unique situation to observe directly the
vertex function in the 
$M_{\pi^0\pi^0}$ spectrum. In other words, the ABC effect in this context is
nothing else but a consequence of the $d^* \to \Delta\Delta$ vertex function
causing  the observed low-mass enhancement by suppression of the high-mass
region. 

We note in passing that already in the work of K\"albermann and Eisenberg
\cite{kaelb} about the ABC effect formfactors had been introduced, but there
at the $\pi NN$ vertices.

In Ref. \cite{prl2011} the cutoff parameter $\Lambda$ had been fitted to the
data in the $M_{\pi^0\pi^0}^2$ spectrum (Dalitz plot projection) resulting in
$\Lambda \approx$ 0.16 GeV/c, which corresponds to a length scale of $r =
\frac{\hbar \sqrt 6}{\Lambda} \approx$ 2 fm. Naively this could be associated
with a hadronic size of $d^*$. However, with a binding of 80 MeV the size of a
tightly bound $\Delta\Delta$ system is expected to be considerably smaller. A
possible solution of this paradox is that this tightly bound $\Delta\Delta$ state
resonates with respect to the $D_{12}\pi$ configuration, which has a much
lower threshold \cite{Avraham}. The $D_{12}\pi$ scenario will be discussed 
further below. 

Fig.~\ref{fig3} shows the effect of the vertex function on the $M_{\pi^0\pi^0}$
spectrum for cutoff parameters $\Lambda$ = 0.1, 0.15, 0.2, 0.3 and 0.6
GeV/c. As we see, values in the range 0.15 - 0.20 GeV/c are preferred by the
data. 

Though this ansatz accounts very well for the data on the double-pionic fusion
reactions to the deuteron (and also to $^3$He and $^4$He \cite{3he,MB,4he,SK},
which are 
not considered here), there are two disturbing features at a first glance:

       First of all, the cutoff parameter $\Lambda$ appears to be unusually
       small. Cutoff parameters used in the description of hadronic reactions
       are usually three to four times larger. However, these are
       conventionally employed for $t$-channel exchanges, see {\it e.g.}
       Refs. \cite{Oset,Zou}. In our case of an
       $s$-channel resonance decay the assumed vertex function is actually
       identical in 
       structure to that appearing in eq. (6) for the $\Delta \to N\pi$
       decay. In fact, there we have $R^{-1}$ = 0.16 GeV/c, which is just
       the value of $\Lambda$ found here for the dibaryon
       resonance. Indeed, such vertex functions are commonly used for the
       description of the decay of baryon  resonances, see {\it e.g.}
       Ref. \cite{Dimitriev,Teis}, with cutoff parameters being similarly small as the
       value obtained here.  

Second, for non-fusion reactions like $pn \to pn \pi^0\pi^0$ the vertex
function in eq. (16) does not affect primarily the $M_{\pi^0\pi^0}$ spectrum,
but the $M_{pn}$ spectrum, as discussed in Ref.~\cite{np00}, since the unbound
$pn$ system is no longer restricted in its relative 
motion by the deuteron wave function. The fact that the vertex function does
not influence significantly the  $M_{\pi^0\pi^0}$ spectrum actually agrees
perfectly with the
experimental finding that there is no significant ABC effect in this reaction
\cite{np00}. 

However, the impact of the vertex function on the $M_{pn}$ spectrum
has been demonstrated in Ref.~\cite{np00} to exaggerate the enhancement of the
low-mass region in the $M_{pn}$ 
spectrum tending thus to be at variance with the data \cite{np00} -- see
Fig.~7 in Ref.~\cite{np00}. 

     In this work the resonance contribution to the $np \to np\pi^0\pi^0$
     reaction was  
     calculated in analogy to that for the $np \to d\pi^0\pi^0$
     reaction, only that three-body phase-space is replaced by a four-body
     phase space and the condition for the momenta of the emitted nucleons
     to be in accord with the deuteron wave function is replaced by the 
     requirement that the emerging $np$ pair undergoes a final-state
     interaction chosen to be of Migdal-Watson type
     \cite{Migdal,Watson,SZC}:
      
\begin{equation}
FSI = 1+\frac{R^{-2}}{(-\frac 1 {a_s} +\frac 12
  r_0 p_{np}^2)^2+p_{np}^2},
\end{equation}

which multiplies the expression for the cross section. Here R denotes the vertex
size, $r_0$ the effective range and $a_s$ the isoscalar triplet scattering
length. For these the values 2.75~fm, 1.75~fm and +5.4~fm have been
used. Whereas the latter two conform to the experimentally known values for
effective range and scattering length, the value of R has been taken to be
identical to the one used in $pp$ induced two-pion production \cite{deldel}.

\begin{figure}
\centering
\includegraphics[width=0.8\columnwidth]{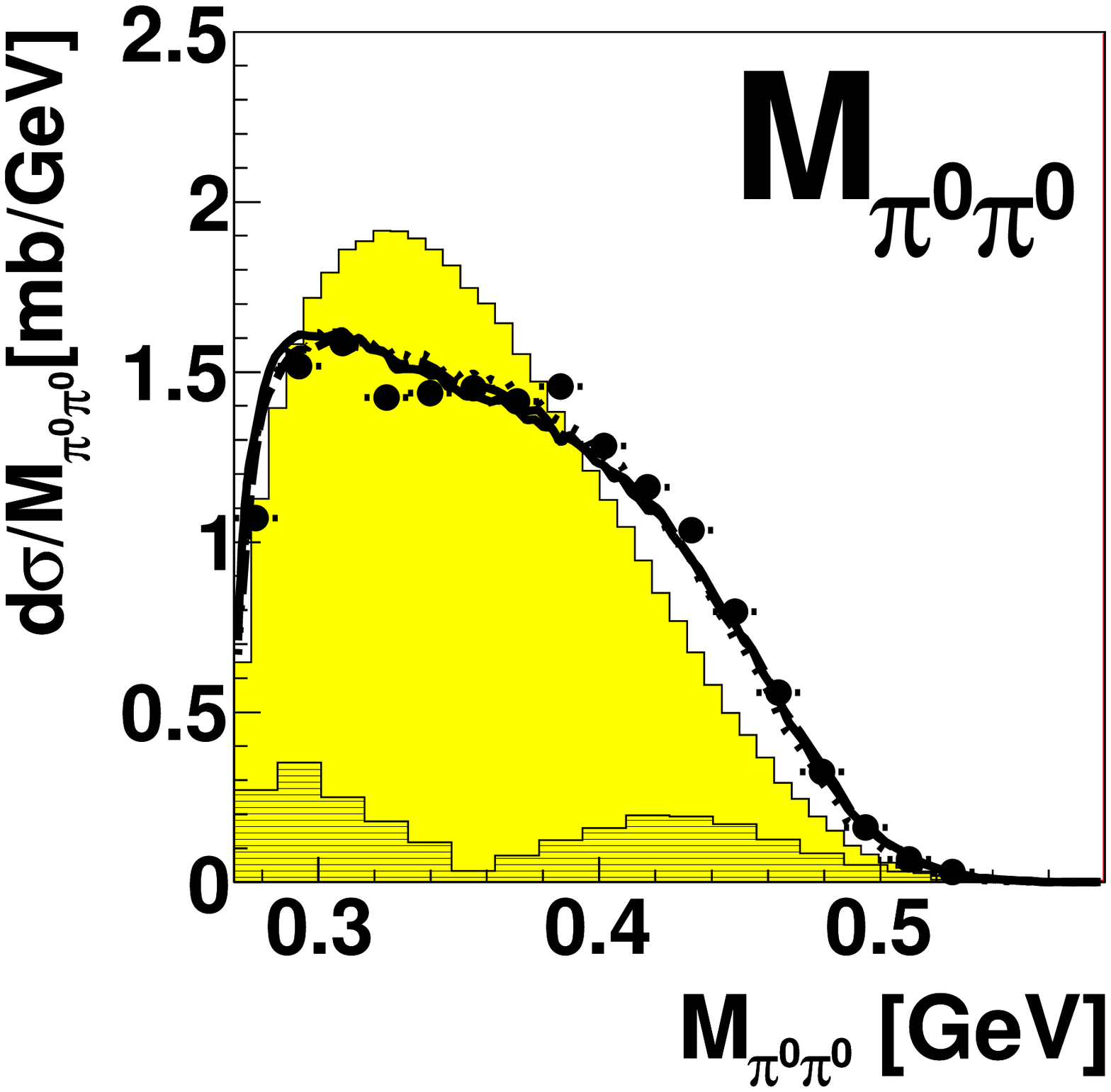}
\includegraphics[width=0.8\columnwidth]{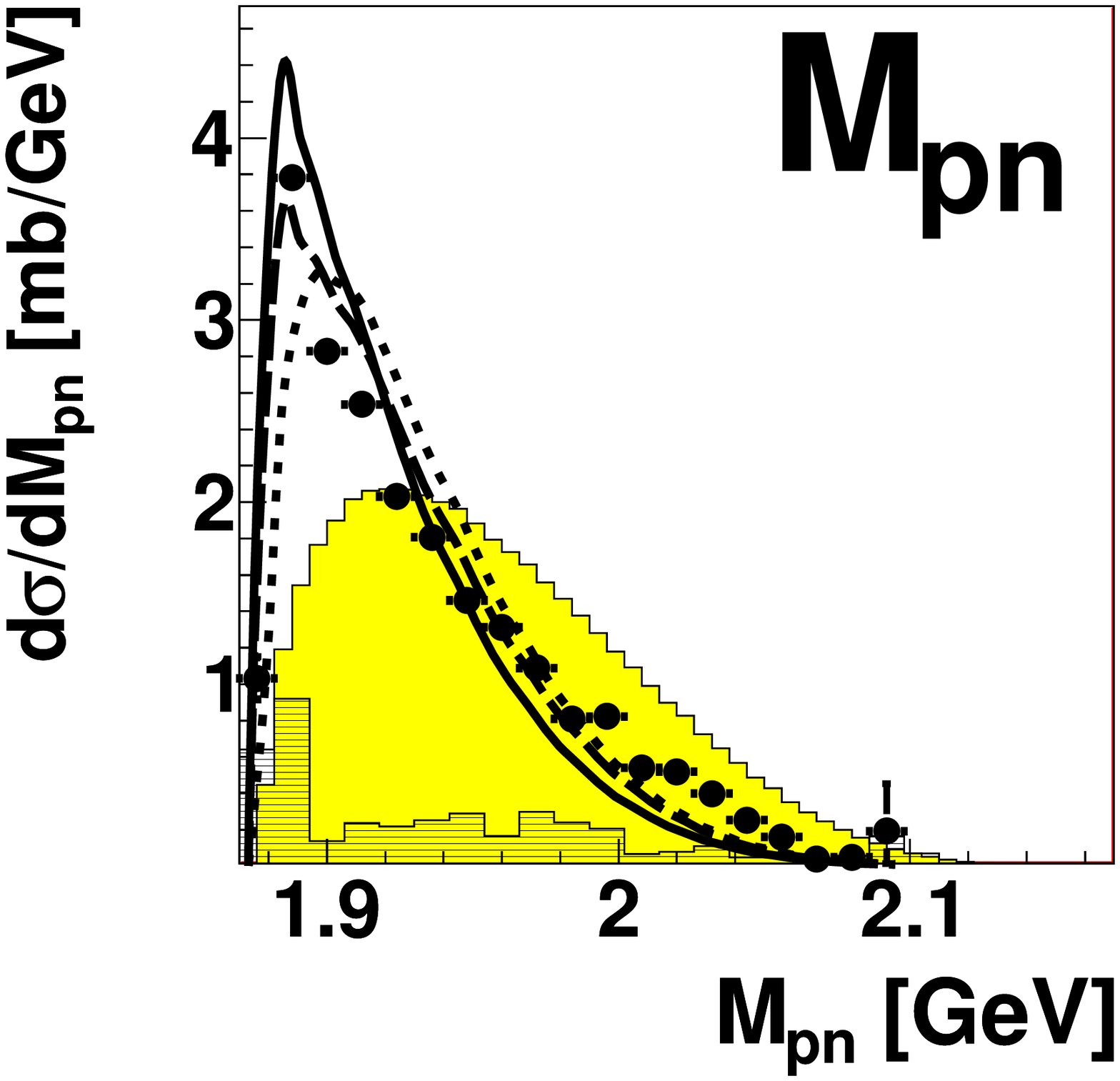}
\caption{\small (color online).
Differential distributions of invariant masses $M_{\pi^0\pi^0}$ (top) and
$M_{pn}$ (bottom) for the $np \to np\pi^0\pi^0$ reaction. The black solid
circles denote data \cite{np00}, the shaded areas indicate the estimate of
systematic uncertainties and the yellow shaded areas phase-space
distributions. The solid (dashed) lines give the calculation of the reaction
process including the route $np \to d^*(2380) \to \Delta\Delta \to
np\pi^0\pi^0$ with (without) use of the vertex function in eq. (16). The
dotted line is a calculation without both vertex function and FSI. Note that
in the $M_{\pi^0\pi^0}$ spectrum all lines lie nearly on top of each
other. 
}
\label{fig4}
\end{figure}

     Re-inspecting that calculation ~\cite{np00} in the course of this work we
     have noted that by mistake the vertex function was applied twice
     there. The updated calculation is plotted in Fig.~\ref{fig4} by the solid
     curve. The dotted curve 
     gives the calculation without FSI and without vertex function, the dashed
     curve shows the calculation with FSI, but without vertex function. We
     still have the situation that the calculation without vertex function
     appears to be slightly superior to that with vertex function. However,
     accounting for the substantial systematic uncertainties in the
     experimental $M_{pn}$ spectrum (hatched areas in Fig.~\ref{fig4}) the
     calculation 
     with vertex function is appropriate as well. We also note that for the
     FSI in these calculations it has been assumed that all emitted $pn$ pairs
     are in relative  $s$-waves, which is supported by the observation, that
     the $M_{pn}$ spectrum cumulates at small invariant masses. 

We just note in passing that -- different from the $np \to d\pi^0\pi^0$ reaction
-- the $np \to np\pi^0\pi^0$ reaction 
contains also isovector contributions from $t$-channel background
processes. In principle one might ask, whether they could possibly be the
cause of the  
different behavior of the $M_{\pi^0\pi^0}$ spectra in both reactions. In the $d^*$
resonance region the background processes make up about thirty percent of the
total cross section. Isospin decomposition according to
Refs. \cite{dakhno,bystricky,iso} gives roughly half of the background
processes to be isovector, {\it i.e.} about 15$\%$ of the total cross
section. Since isoscalar and isovector contributions do not interfere in total
cross section and invariant-mass spectra, the influence of the isovector part
there is small. Moreover, $d^*$ and $t$-channel background processes give nearly
identical shapes for the $M_{\pi^0\pi^0}$ spectrum -- see Fig.~7 in
Ref. \cite{np00}.

In the following we examine three alternative ansatzes, which have been
given recently in the literature for the explanation of the ABC
effect. We also discuss as a new aspect the 
possibility of a $D$-state between the $\Delta$s in the intermediate
$\Delta\Delta$ system.  

We do not deal here with the model of Gardestig, F\"aldt and Wilkin \cite{GFW},
since this has 
been designed specifically for the ABC effect in the double-pionic fusion to
$^4$He and is not applicable to the basic double-pionic fusion to the
deuteron.

\subsection{$\Delta\Delta$ Final State Interaction}

 Another, alternative attempt to explain the ABC-effect by a final-state
 interaction (FSI)  
 between the two $\Delta$s in the intermediate state dates back to the first
 WASA measurements on this issue \cite{MB,MBthesis}. We note that this ansatz
 dates back to the time, when it was not yet known that the ABC effect is
 correlated with the presence of the $s$-channel resonance $d^*(2380)$. Though
 including a FSI for ejectiles, which beforehand formed a resonance, has the
 threaten of double-counting, we will discuss this scenario for historical
 reasons - and also because its effect is formally very similar to that of the
 vertex function.

\begin{figure}
\centering
\includegraphics[width=0.99\columnwidth]{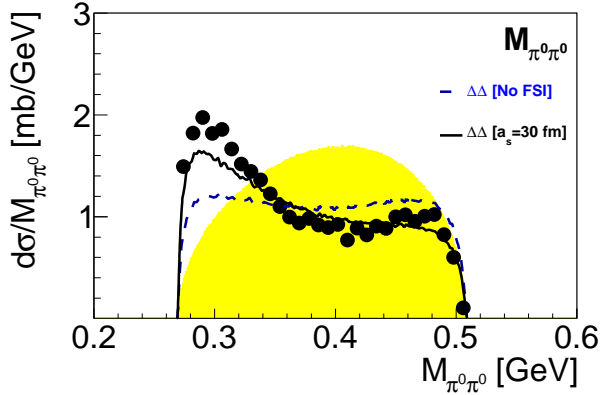}
\caption{\small (color online). The same as Fig.~\ref{fig3}, but the lines represent
  now the effect of the $\Delta\Delta$-FSI in dependence of the scattering
  length. 
}
\label{fig5}
\end{figure}

           In Refs. \cite{MB,MBthesis} the $\Delta\Delta$-FSI was parameterized in the
           Migdal-Watson ansatz \cite{Migdal,Watson,SZC} providing the factor
-- analogous to eq. (17) --
\begin{equation}
FSI = 1+\frac{R^{-2}}{(-\frac 1 {a_s} +\frac 12
  r_0 p_{\Delta\Delta}^2)^2+p_{\Delta\Delta}^2},
\end{equation}

which multiplies the expression for the cross section. 
For decent
values of $r_0$~=~2~fm and $R$ = 1.1 fm a reasonable description of the
$M_{\pi^0\pi^0}^2$ spectrum is achieved with $a_s \approx$ 30 fm, see
Fig.~\ref{fig5}. Numerically the first term in eq. (18) is small compared to
the second 
one and also the finite range term in the denominator of the second term may
be neglected. Thus we see that  eqs. (16) and (18) both have a similar
structure with similar $p_{\Delta\Delta}$ dependence, {\it i.e.} they lead to
practically the same corrections in the differential spectra. 


Thus eq. (18) formally 
leads back to eq. (16) for the vertex function. 
Actually, this ansatz seems to be independent of the existence of the $d^*$
resonance and hence also valid for $t$-channel $\Delta\Delta$ excitations,
where no ABC enhancements are observed. However, the scattering length in
eq. (18) depends on the involved partial wave with isospin I and total spin
J. The fact that the deuteron angular distribution gets flatter again above
the $d^*$ region \cite{pol00} in the $t$-channel $\Delta\Delta$ regime means
that there are different partial waves involved, which necessarily do not need
to have a large scattering length.


\subsection{ $d^*(2380) \to d\sigma$ decay}

Recently Platonova and Kukulin proposed an alternative explanation of
the ABC effect \cite{Kuk,Kukulin}. They consider two possible decay branches for
the $d^*(2380)$ resonance. Hereby they go back to the work of Dyson and
Xuong \cite{Dyson}, who -- based on SU(6) -- predicted a sextet of non-strange
dibaryon states. In that work $d^*(2380)$ is denoted by $D_{03}$, where the
first index means the isospin and the second one the spin of the dibaryon
state.  

\begin{figure}
\centering
\includegraphics[width=0.99\columnwidth]{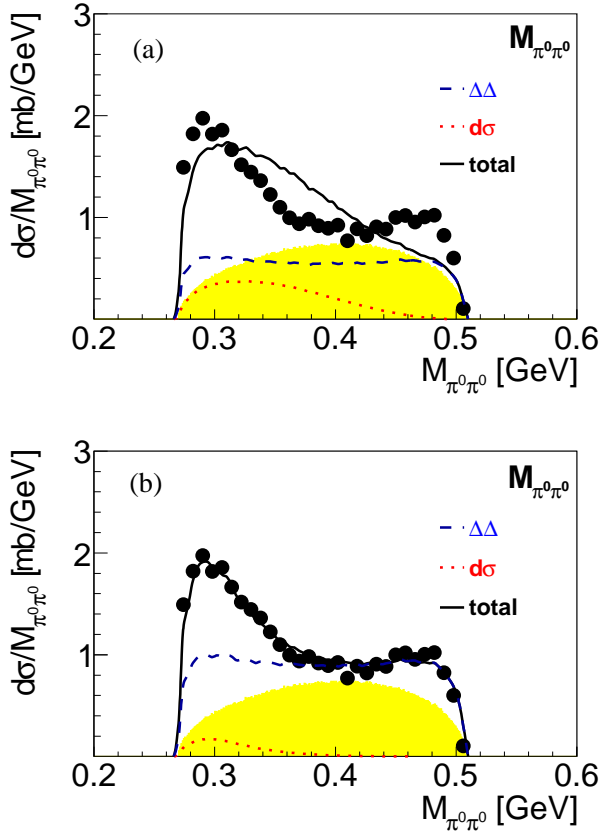}
\caption{\small 
(color online). 
  The same as Fig.~\ref{fig3}, but the lines represent now the model ansatz of Kukulin
  and Platonova \cite{Kuk} for the $d^*(2380)$ decay. The contribution of
  the main route $d^*(2380)\to D_{12}\pi^0 \to d\pi^0\pi^0$ is given by the blue
  dashed line, which to very good approximation is represented by a
  $\Delta\Delta$ calculation without vertex function. The red dotted lines
  show the contribution of the route $d^*(2380) \to d\sigma \to d\pi^0\pi^0$
  -- in the top panel (a) for mass and width 
  of the $\sigma$ meson according to PDG \cite{PDG} ($m_\sigma$ = 440 MeV,
  $\Gamma_\sigma$ = 544 MeV), and in the bottom panel (b) for the
  values $m_{\sigma} = 300$ MeV and $\Gamma_{\sigma}$ = 100 MeV as used in
  Ref.~\cite{Kuk}. The solid lines give the coherent sum.
}
\label{fig6}
\end{figure}

Kukulin and Platonova assume the main decay of $D_{03}$ to proceed via the
$D_{12}$ member of the sextet, {\it i.e.} $D_{03}\to D_{12} \pi \to d\pi\pi$, 
where $D_{12}$ is identified with the $I(J^P) = 1(2^+)$ state at the $N\Delta$
threshold, which produces a pole in the $^1D_2$ partial wave of $pp$
scattering. For the numerous discussions, whether this state constitutes a 
true $s$-channel resonance or not, see {\it e.g.}
Refs. \cite{said0,said1,said2,said3,shypit1,shypit2,sarantsev,strakovsky,igor,hos1,hos2,Seth}.   
Since in this route $D_{12}$
and the associated pion have to be in relative $p$-wave, in order to fulfill
the angular momentum requirements of $D_{03}$, this decay route is practically
indistinguishable from the route $D_{03} \to \Delta\Delta \to d\pi\pi$, {\it
  i.e.} both routes give essentially identical results for the $M_{\pi^0\pi^0}$
spectrum (dashed lines in Fig.~\ref{fig6}). 

Note that in Ref.~\cite{Kuk} momentum
dependent decay widths containing also vertex functions analogous to that in
eq. (16) have been used with very small cutoff parameters $\Lambda$ = 0.1 -
0.2 GeV/c. Since 
the decays $d^* \to D_{12} \pi$ and $D_{12}\to d\pi$ are $p$-wave decays,
which primarily go with the third power of the decay momentum, a vertex
function with very small cutoff parameter tends to cancel this momentum
dependence. Hence the calculation of the decay route $d^* \to D_{12} \pi \to
d\pi\pi$ with very small cutoff parameters gets very similar to that of $d^*
\to \Delta\Delta \to d\pi\pi$ without the use of a vertex function for the
$d^* \Delta\Delta$ decay vertex. 

The second decay route constitutes a really new piece in trying to explain the
ABC effect. It assumes that a $\sigma$ meson is emitted according to 
$D_{03}\to d\sigma$. Due to $J^P = 3^+$ the $\sigma$ meson and the deuteron
have to be in relative $d$-wave. This means that the transition amplitude of
this process should be proportional to the relative momentum squared, {\it
  i.e.} $A \sim q_{\sigma-d}^2 \sim
 (M_{\pi\pi^{max}}^2-M_{\pi\pi}^2)$. This
momentum dependence provides a high-mass $M_{\pi\pi}$ suppression
(dotted line in Fig.~\ref{fig6}(a)). In total, both routes -- if they are
allowed to interfere constructively -- produce 
an enhancement of the low-mass region relative to the high-mass region as
required for the ABC effect (solid line in Fig.~\ref{fig6}(a)). Only, the shape of
the produced low-mass enhancement comes out too broad in comparison to the data.

In this ansatz the data are only reproduced quantitatively, if a light and
narrow $\sigma$ meson ($M_{\sigma}=300 MeV$, $\Gamma_{\sigma}=100 MeV$) is
assumed (solid line in Fig.~\ref{fig6}(b)). The strong mass reduction compared to the
accepted value \cite{PDG} has been assigned to chiral symmetry restoration
inside the $d^*$ dibaryon \cite{Kuk}.

An admixture of this decay route as small as 5$\%$ on the cross section level
is enough to reproduce the 
ABC effect in double-pionic fusion reactions. A problem with this
ansatz may arise for the $np \to np\pi^0\pi^0$ reaction, where the data
exhibit no ABC effect in the $M_{\pi^0\pi^0}$ spectrum. Since in this reaction
an isoscalar pion pair is produced, which can form a $\sigma$ meson, there is no
obvious reason, why the $D_{03}\to pn\sigma$ route should be 
suppressed. However, in Ref. \cite{Kukulin} it is argued qualitatively that
due to the much increased phase space for the unbound $NN$ system the
centrifugal barrier for the $d$-wave emission of the $\sigma$ meson plays here
a much bigger role and might suppress its emission more than in the $d\pi\pi$
case. Unfortunately, a convincing quantitative calculation has not yet been
presented by these authors. 



\subsection{{\it D-wave} $\Delta\Delta$ admixture}

Having learned in the previous example that a $D$-wave introduces a strong
momentum dependence such that the ABC effect may be reproduced, we may
consider yet another ansatz, which possibly is more realistic.

Similar to the fact that the deuteron groundstate contains a $D$-wave
admixture, we may postulate also a such one for the intermediate
$\Delta\Delta$ system. 
In principle a $D$-wave between the two $\Delta$s may couple to spin  $S$ = 1
and 3 states of the $\Delta\Delta$ system for a total angular momentum of
$J$ =3. For simplicity we consider here just the spin $S$~=~3 case as done in
Refs. \cite{zhang1,zhang2}. 

\begin{figure}
\centering
\includegraphics[width=0.99\columnwidth]{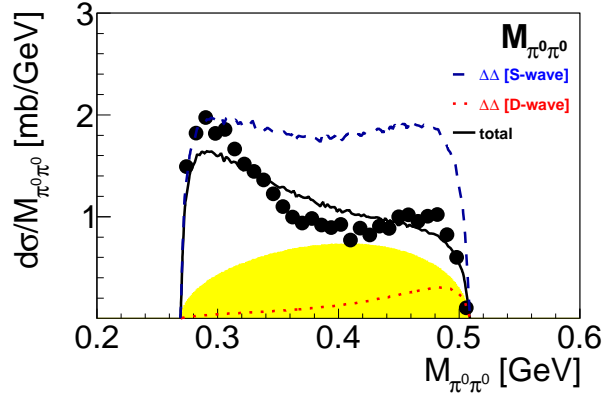}
\caption{\small  (color online).The same as Fig.~\ref{fig3}, but the dotted line
  represents now a 5$\%$ contribution of the $D$-wave in the intermediate
  $\Delta\Delta$ system and the solid curve the coherent sum of $S$- and
  $D$-wave contributions -- without vertex function.
}
\label{fig7}
\end{figure}

\begin{figure}
\centering
\includegraphics[width=0.79\columnwidth]{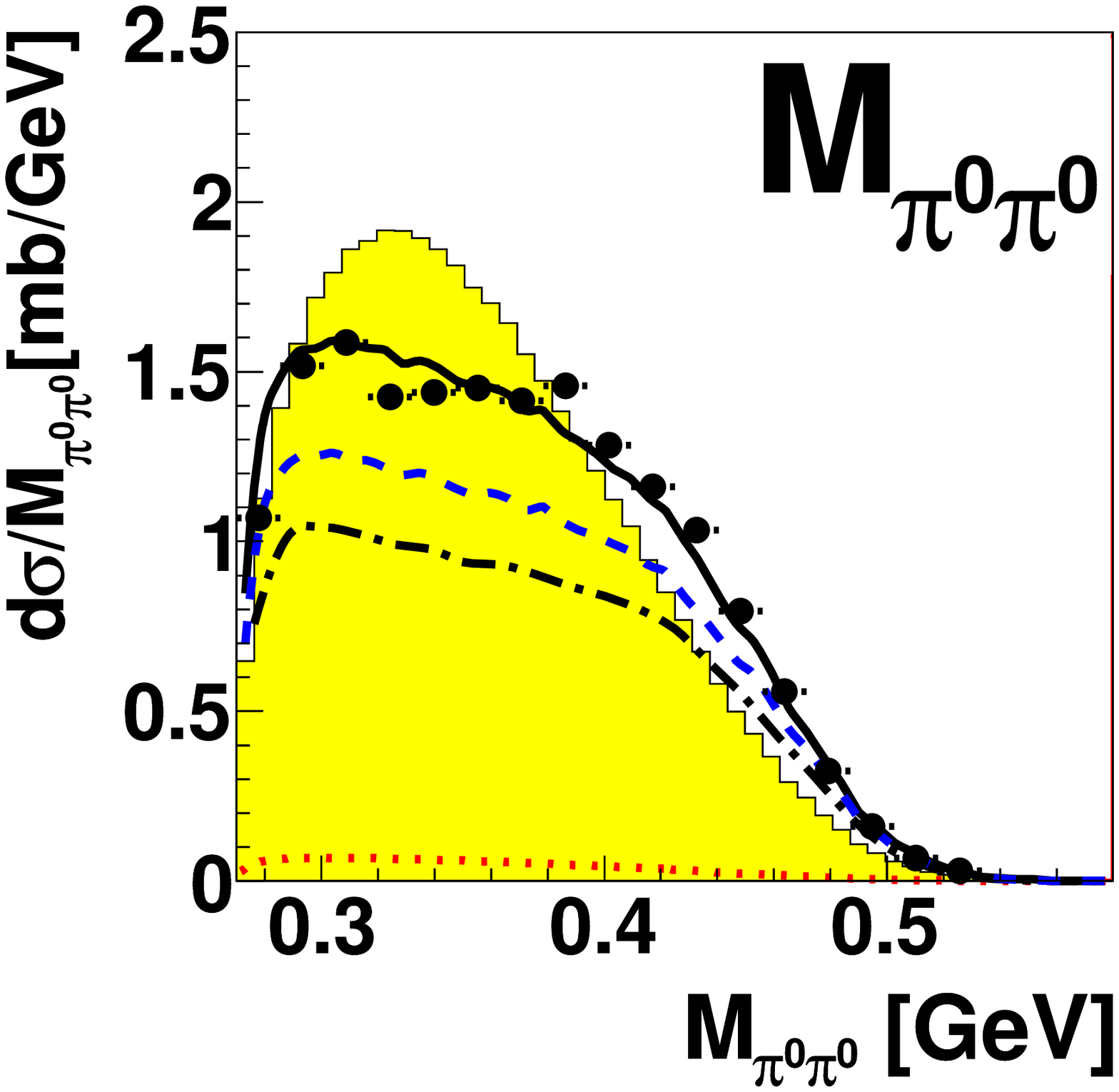}
\includegraphics[width=0.79\columnwidth]{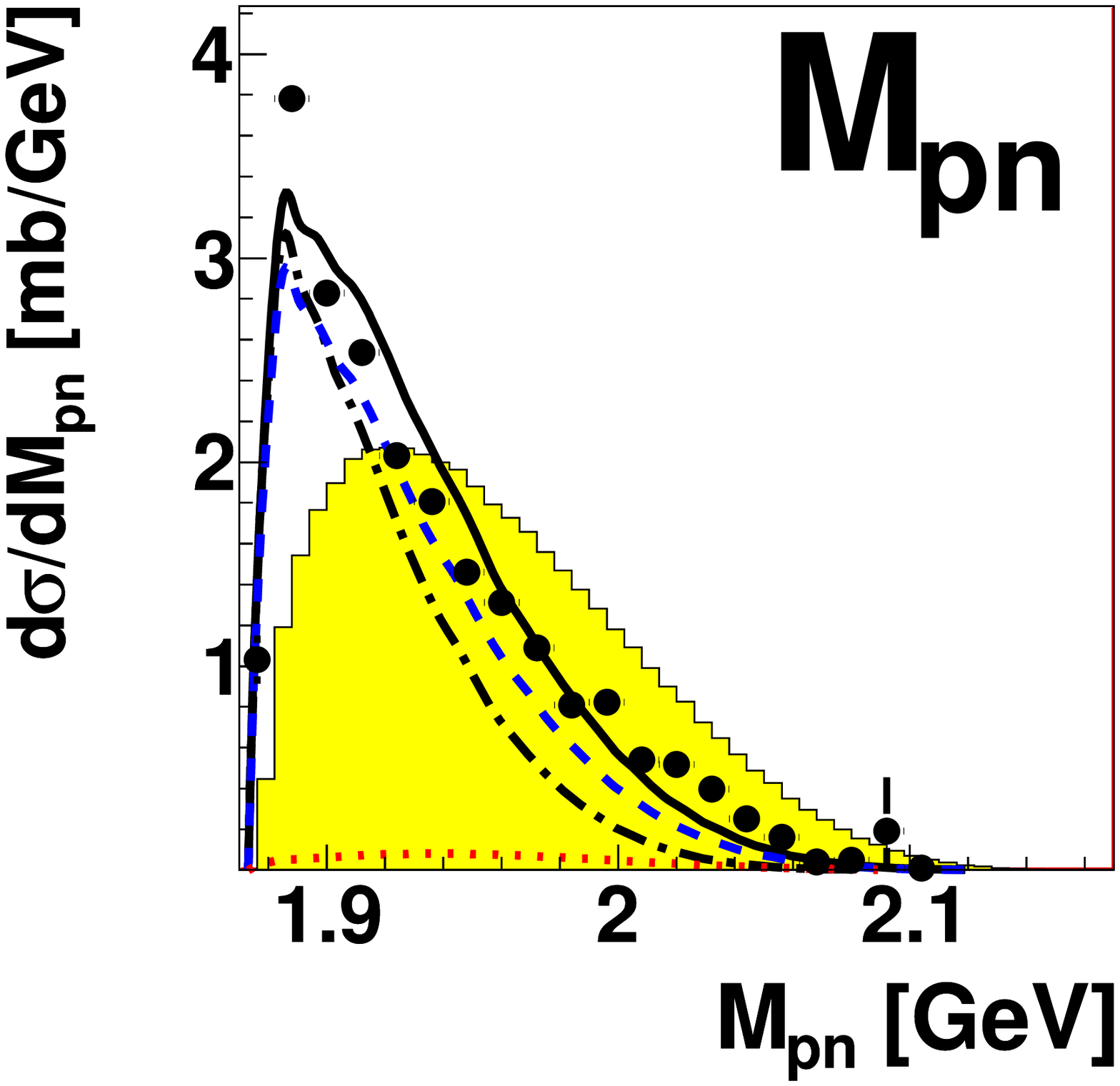}
\caption{\small  (color online). The same as Fig.~\ref{fig4} for the $np \to
  np\pi^0\pi^0$ reaction, but dashed and dotted lines
  represent now the $d^*$ contributions originating from $S$ and $D$-waves in
  the intermediate $\Delta\Delta$ system, respectively. The dash-dotted
  curves show their coherent sum. The solid lines give
  the full result including background contributions.
}
\label{fig8}
\end{figure}

The $D$-wave $\Delta\Delta$ amplitude is
proportional to $q_{\Delta\Delta}^2$. So contrary to the ansatz with the
$\Delta\Delta$ vertex function and $\Delta\Delta$ FSI, respectively, the
$M_{\pi\pi}$ spectrum will be suppressed in particular at small masses, see
dotted line in Fig.~\ref{fig7}. A 
destructive interference between $D$-wave and $S$-wave (dashed line)
components will suppress then high $M_{\pi\pi}$ invariant masses. A 
20$\%$ admixture of the $D$-wave component on the amplitude level, {\it i.e.}
4 - 5$\%$ on the cross section level, is sufficient for a reasonable
description of the ABC-effect. This means a very similar amount of $D$-wave
admixture as in the deuteron groundstate.
 


The angular distributions are not affected noticeably by the
$D$ wave $\Delta\Delta$ admixture, since the deuteron angular distribution is
characterized by the total angular momentum of $d^*$ and the pion angular
distribution by the $p$-wave decay of the intermediate $\Delta$ excitations
\cite{prl2011}. 

Opposite to the previous cases, this ansatz does not cause a problem in the
description of the $np \to np\pi^0\pi^0$ reaction. It does neither produce an
ABC effect in the $M_{\pi^0\pi^0}$ spectrum nor an exaggerated low-mass
enhancement in the $M_{pn}$ spectrum. The situation is depicted in
Fig.~\ref{fig8}, 
where the data from Ref. \cite{np00} are compared with the results of the
ansatz with $S$- (dashed) and $D$-wave (dotted) $\Delta\Delta$
configurations. Whereas these calculations produce practically identical
shapes in the $M_{\pi^0\pi^0}$ spectrum, they differ somewhat for the $M_{pn}$
case. The solid lines give the full result, {\it i.e.} the coherent sum of 
$S$ and $D$-wave contribution together with the $t$-channel background
processes. The data of the $np \to np\pi^0\pi^0$ reaction are well reproduced
by this ansatz.  

As discussed already in connection with the vertex function and in
Ref. \cite{pp0-} -- the $np$ final system in the unbound case is no longer
restricted by the deuteron wave function. Different from the bound nucleus
case, where the relative momentum between the two $\Delta$s is essentially
made up by the relative momentum between the two emerging pions, the relative
$\Delta\Delta$ momentum in the unbound case is mainly transferred to the two
emerging nucleons, the heavy partners of the $\Delta$ decays. Hence the
introduction of $q_{\Delta\Delta}^2$ has no visible influence on the shape of
the $M_{\pi^0\pi^0}$ spectrum, but rather on the $M_{pn}$ spectrum -- see
Fig.~\ref{fig8}, bottom.   

\begin{figure}
\centering
\includegraphics[width=0.99\columnwidth]{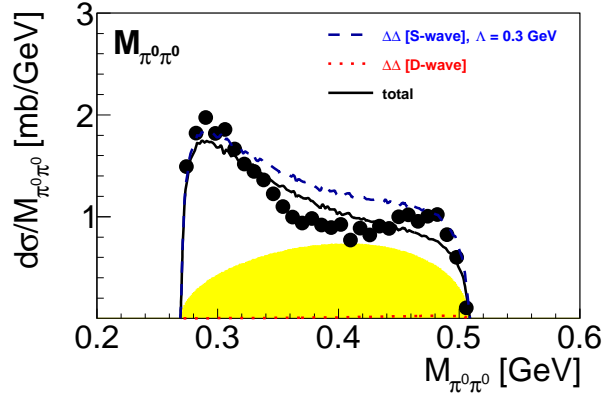}
\caption{\small  (color online).The same as Fig.~\ref{fig7}, but the dotted line
  represents now a only 1.5$\%$ contribution of the $D$-wave in the intermediate
  $\Delta\Delta$ system and the solid curve the coherent sum of $S$- and
  $D$-wave contributions incorporating now also a vertex function with
  $\Lambda$ = 0.3 GeV/c. 
}
\label{fig9}
\end{figure}

A $\Delta\Delta$ $D$-wave contribution has actually been predicted in 
quark model calculations recently \cite{zhang1,zhang2}. The predicted
$d$-wave percentage, however, is only 1.5$\%$. In fact, we may also reproduce
the ABC effect with such  a small $D$-wave admixture, if we incorporate also
the vertex function in this ansatz and increase  the cutoff parameter to
$\Lambda$ = 0.3 GeV/c -- see Fig.~\ref{fig9}. That way we may arrive at a 
consistent description of the ABC effect, which leads also to an optimal
description of both $np \to d\pi^0\pi^0$ and $np \to np\pi^0\pi^0$ reactions
-- though the interplay between cutoff parameter and $D$-wave contribution
allows quite some freedom, as long as one of them is not fixed by other means.

\subsection{$d^*(2380) \to N^*(1440)N$ decay}

This scenario has been
considered to some extent in Ref. \cite{Bugg}. 
Since the $d^*(2380)\to NN$ decay is sizable \cite{npfull} and the Roper
resonance $N^*(1440)$ has the same quantum numbers as the nucleon, one may
assume that a $d^*(2380)\to N^*(1440)N$ decay branch might also exist. Due to
the spin-parity $J^P 
= 3^+$ of the $d^*(2380)$ resonance one needs to have at least a $D$-wave
between the nucleon and the Roper resonance, which in the course of its decays
$N^*(1440)\to N\sigma \to N\pi\pi$ and $N^*(1440)\to \Delta\pi \to N\pi\pi$
finally must transform to a $D$-wave between pion and nucleon pairs, if the
nucleons fuse to a deuteron. This affords substantial reshuffling of angular
momenta.

\begin{figure}
\centering
\includegraphics[width=0.99\columnwidth]{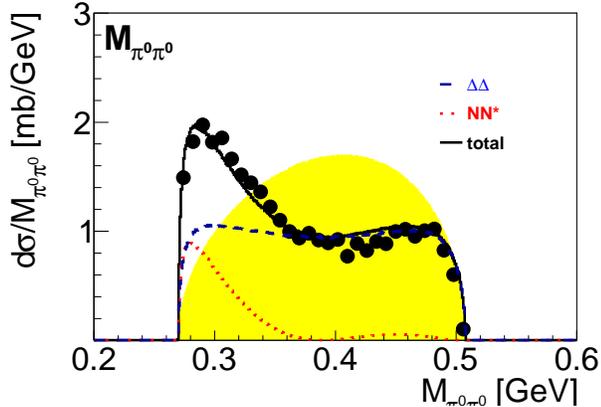}
\caption{\small (color online).
   Same as Fig.~\ref{fig3}, but dashed (blue)1
  and dotted (red) lines show $d^* \to \Delta\Delta \to d\pi\pi$ and
  $d^* \to N^*N \to \Delta \pi N \to d\pi\pi$ contributions
  respectively -- without applying a vertex function at the decay vertex. The
  solid line represents their coherent sum. 
}
\label{fig10}
\end{figure}

Also since the mass of $d^*(2380)$ is just at the $N^*(1440)N$ threshold, we
can not expect that the amplitude for a decay into a intermediate $N^*(1440)N$
system in relative $D$-wave is large. But let us assume that it exists in
some sizable amount.  Then it will interfere with the main decay branch
$d^*(2380)\to\Delta\Delta$. In fact, assuming a 5$\%$ decay branch via the
Roper resonance the ABC-effect can be again very well reproduced
     -- see Fig.~\ref{fig10}.

Since in the $np \to np\pi^0\pi^0$ reaction the emitted $np$ pair is no longer
forced to be in 
relative $S$ wave, the $d^* \to NN^*$ decay even could be enhanced in this
channel. The effect of this decay branch in $\pi^0\pi^0$- and $pn$-invariant
mass spectra is similar to that discussed for the
$D$-wave $\Delta\Delta$ scenario.

If true that decays of baryonic excitations should have a vertex function in
general, 
then also in this scenario the main decay route via $\Delta\Delta$ will
already produce the main part of the ABC effect with the consequence that the
necessary amount of $d^* \to NN^*$ decay is diminished to a small
percentage similar to the situation discussed for the $D$-wave $\Delta\Delta$
scenario above.

%
%
%
If this scenario is correct then we expect also a decay of
$d^*(2380)$ into the $NN\pi$ channel, since the decay of the Roper resonance
into $N\pi$ has a branching of 55 - 75$\%$~\cite{PDG}.

\section{Conclusions}

We have discussed several possible reasons for the ABC effect. Most of these
hypotheses can be tuned to reproduce at least qualitatively the measured
low-mass enhancements in the $M_{\pi\pi}$ spectra of double-pionic fusion
reactions. The non-occurrence of the ABC effect in the non-fusion
reaction $np \to np\pi^0\pi^0$ has been used as a further constraint to
filter out viable solutions. 

Most natural and straightforward appears the explanation of the ABC effect as
a direct consequence of the vertex function in the $d^* \to \Delta\Delta$
decay vertex. In this conception the $M_{\pi\pi}$ spectrum just maps the
momentum dependence of the resonance decay, which is given by that of the
vertex function. The necessary cutoff parameter happens to coincide with that
for the $\Delta$ resonance and is in the bulk part of values used for other
baryonic excitations. This scenario also 
naturally explains the non-appearance of the ABC effect in non-fusion
reactions. There remains some
tension in the description of the $M_{pn}$ spectrum in the $np \to
np\pi^0\pi^0$ reaction. However, this is still within the systematic
uncertainties of the particular measurement. 

Alternative hypotheses deal with postulated small $D$-wave admixtures either
in the  intermediate $\Delta\Delta$ system or in form of $d\sigma$ or
$N^*(1440)N$ decay branches. 
A $D$-wave admixture in the  intermediate $\Delta\Delta$ system has been
predicted from quark-model calculations \cite{zhang1,zhang2} -- though a
factor of three less than needed for an exclusive explanation of the ABC
effect without the use of the vertex function. However, if also in
this scenario a vertex function is incorporated -- as needed for a 
consistent description of baryonic excitations, then a 1.5$\%$
$d$-wave admixture in agreement with model predictions is sufficient for a
quantitative description of both key reactions. 


With regard to a possible $d^*(2380) \to N^*(1440)N$  decay a dedicated
experimental test of such a decay could be the 
investigation of isoscalar single-pion production in the energy region of
interest. In addition, investigation of such a decay branch into the $NN\pi$
system can resolve the question, whether the main decay route is $d^*(2380)
\to \Delta\Delta$ or $d^*(2380) \to D_{12}\pi$ as anticipated in
Refs. \cite{GG1,GG2,Kuk}. Since $D_{12}$ has a sizeable decay branch into
$NN$, the route via $D_{12}$ will feed also the $NN\pi$ channel, whereas such
a feeding is 
     very unlikely to happen with sizeable intensity in case of
 an intermediate isoscalar $\Delta\Delta$ system.

Unfortunately, there are no data at all in this energy region
and even at lower energies the data base is not very precise
\cite{isosarantsev}. However, 
though WASA at COSY has finished data taking, there are still data samples from
previous runs available, which -- though collected primarily for other reasons
-- could allow to extract the wanted isoscalar single-pion production. An
analysis of such data is in progress.

Thus the ABC effect in double-pionic fusion reactions together with its
absence in non-fusion reactions may lead to some
insight into the internal structure of the $d^*(2380)$ resonance. This is also
an important aspect in the photo excitation $\gamma d \to d^*(2380)$. By use
of virtual $\gamma$s even the transition formfactor may be obtained, which
contains information about size and internal structure of this 
resonance state. Dedicated experiments with real $\gamma$s are expected to be
conducted at MAMI in near future.

%
%

\section{Acknowledgments}

We acknowledge valuable discussions with V. Baru, A. Gal, V. Grishina,
C. Hanhart, G. K\"albermann, 
V. Kukulin, E. Oset, G. J. Wagner and C. Wilkin. This work has been supported
by 
DFG (CL 214/3-1) and STFC (ST/L00478X/1).

\end{document}